\documentclass[twocolumn,amssymb,superscriptaddress,prl,showpacs]{revtex4}
\usepackage{graphicx}
\usepackage{dcolumn}
\usepackage{bm}

\begin{document}

\title{Violation of the Wiedemann-Franz Law in a Single-Electron Transistor}

\author{Bj\"orn Kubala}
\author{J\"urgen K\"onig}
\affiliation{Institut f\"ur Theoretische Physik III,
Ruhr-Universit\"at Bochum, 44780 Bochum, Germany }

\author{Jukka Pekola}
\affiliation{Low Temperature Laboratory, Helsinki University of Technology,
PO BOX 3500, 02015 TKK, Finland}

\begin{abstract}
We study the influence of Coulomb interaction on the thermoelectric transport
coefficients for a metallic single-electron transistor.
By performing a perturbation expansion up to second order in the tunnel-barrier
conductance, we include sequential and cotunneling processes as well as
quantum fluctuations that renormalize the charging energy and the tunnel
conductance.
We find that Coulomb interaction leads to a strong violation of the Wiedemann-Franz law: the Lorenz ratio
becomes gate-voltage dependent for sequential tunneling, and is increased by a
factor $9/5$ in the cotunneling regime.
Finally, we suggest a measurement scheme for an experimental realization.
\end{abstract}



\date{Received 26 September 2007; published 14 February 2008}

\maketitle

{\it Introduction.} --
Electron transport in conductors is accompanied with the transfer of both
charge and heat (energy).
Thermoelectric transport coefficients relate the charge and
heat current, $I_e$ and $I_q$, to applied voltage and temperature
differences, $\Delta V$ and $\Delta T$,
\begin{equation}
  \left(\begin{array}{c} I_e \\ I_q \\ \end{array}\right)
  =
  \left(\begin{array}{cc} G_V &  G_T \\ M & K \\
  \end{array}\right)
  \left(\begin{array}{c} \Delta V \\ \Delta T\\ \end{array}\right) \,.
\end{equation}
The thermal conductance $\kappa$ is defined by $I_q = \kappa \Delta T$ for
 $I_e=0$, i.e., $\kappa = K - G_V T S^2$ where
$S = - \Delta V / \Delta T = G_T / G_V$ denotes the thermopower.
For macroscopic samples of ordinary metals, the Wiedemann-Franz law provides
a universal relation between the two conductances by stating that the
Lorenz ratio
\begin{equation}
  L \equiv \frac{\kappa}{G_V T} \, ,
\end{equation}
is a constant given by the Lorenz number $L_0 = (\pi^2/3) (k_B / e)^2$.
It is a consequence of Fermi-liquid theory, which is applicable when
screening renders Coulomb interactions sufficiently weak.
The Wiedemann-Franz law indicates that both charge and heat currents are
supported by the same underlying scattering mechanisms with only weak
energy dependence.

The situation is fundamentally different in mesoscopic systems in which
level quantization and Coulomb interaction drastically affects transport.
The thermopower has been measured in small dots with discrete level spectrum
\cite{discrete-dots-exp}, chaotic dots \cite{chaotic-dots-exp},
carbon nanotubes and molecules \cite{SmallPRL03}, and dots closer to the metallic
(quasi-continuous) limit \cite{StaringEPL93}, and calculated for various
mesoscopic systems \cite{thermopower-theory,kubala06}.
Deviations from the Wiedemann-Franz law have been predicted for
tunneling transport through quantum dots for weak 
coupling \cite{BoeseEPL01,tsaousidou}, in the Kondo 
regime \cite{BoeseEPL01,Kondo-dot}, for open dots \cite{open-dot}, and for granular metals \cite{beloborodov05}.

In the present Letter we address the question of whether and how Coulomb
interaction in a metallic single-electron transistor (SET) with weak tunnel couplings
affects the Wiedemann-Franz law.
The Coulomb interaction plays an important role for two reasons.
First, the finite charging energy to add or remove an electron to or from
the island suppresses some transport processes.
This dramatically affects the charge and thermal conductance
individually, but leaves Wiedemann-Franz law untouched
since the same transport processes are suppressed for both electric and
thermal conductance.
Second, however, Coulomb interaction leads to a strong energy dependence
of the scattering processes.
This yields, in general, a violation of the Wiedemann-Franz law.

By performing a systematic perturbation expansion of all thermoelectric
coefficients up to second order in the tunneling conductances, we calculate the
effect of Coulomb interaction on the Lorenz ratio.
We find that the latter is increased due to Coulomb interaction. 
In the low-temperature regime, the Lorenz ratio becomes gate-voltage
dependent.
For the sequential-tunneling contributions it remains $L_0$ only exactly on
the resonance points where the charging-energy gap of the relevant
transport process vanishes.
With increased detuning the Lorenz ratio rises quadratically as a function
of the charging-energy gap.
Interestingly, we find that in the cotunneling regime, the Lorenz ratio
becomes universal again, but by a factor $9/5$ larger than $L_0$.

{\it Model.} --
We model the metallic single-electron transistor by the Hamiltonian
$H=H_L+H_R+H_I+H_{\rm ch}+H_T$.
Here, $H_{r}=\sum_{kn}\epsilon^{r}_{kn}a^\dagger_{{r}kn} a_{{r}kn}$ with
$r=L,R$ and $H_{I}=\sum_{qn}\epsilon_{qn} c^\dagger_{qn} c_{qn}$
describe noninteracting electrons in the left and right lead and on the island,
respectively \cite{many_channel}. 
Coulomb interaction on the island is described by
the capacitance model $H_{\rm ch}=E_C(\hat{N}-n_x)^2$,
where $E_C=e^2/(2C)$ defines the charging-energy scale with total island
capacitance $C=C_L+C_R+C_g$, $\hat{N}$ is the number operator of excess
charges on the island, and $en_x=C_L V_L+C_R V_R+C_g V_g$ is the
``external charge'' that is tunable by gate and bias voltage.
To increase the number of electrons on the island from $N$ to $N+1$ one has to
overcome the charging-energy gap
 $\Delta_N = \left\langle N+1|H_{\text{ch}}
|N+1 \right\rangle -  \left\langle N|H_{\text{ch}}|N \right\rangle = E_C
\left[ 1+2\left(N-n_x\right)\right]$.
The resonance condition $\Delta_N =0$, where the charging-energy gap vanishes,
is fulfilled at half-integer values of $n_x$.
Charge transfer processes are described by the tunneling Hamiltonian
$H_{T}=\sum_{r=L,R}\sum_{kqn} T^{r} a^\dagger_{rkn}c_{qn}
{e}^{-i\hat{\varphi}}+ {\rm h.c.}$, where
we can assume the tunnel matrix elements
$T^r$ to be independent of the states $k$ and $q$ and channel index $n$, as they vary on the scale of the Fermi energy, which is much larger than all other relevant energy scales.

The tunnel-coupling strength for barrier $r=L,R$ is characterized by the
dimensionless conductance $\alpha_0^r = \sum_n N_{r}(0) N_I(0)|T^{r}|^2 \ll 1$,
where $N_{I/r}(0)$ are the density of states of the island/leads at the Fermi
level, and $\alpha_0 \equiv \sum_{r=L,R}\alpha_0^r$.
The operator ${e}^{\pm i\hat{\varphi}}$ shifts the charge on the island by
$\pm e$.
In general, the electron temperatures of left lead, island, and right lead can
all be different from each other and differ from the lattice temperature.

{\it Thermoelectric coefficients.} --
Charge current $I^{r}_e= e\langle \sum_{kn} dN_{rkn}/dt\rangle$ and heat current $I^{r}_{q}=-\langle \sum_{kn}(\epsilon_{kr}^r-\mu_r) dN_{rkn}/dt\rangle$ leaving lead $r=L,R$ are given by
\begin{eqnarray}
    I^{r}_e &=& -\frac{ie}{\hbar} \int d\omega
    \left[ \alpha^{r+}_e(\omega) C^>(\omega) + \alpha^{r-}_e(\omega)
      C^<(\omega) \right]
    \\
    I^{r}_{q} &=& \frac{i}{\hbar} \int d\omega
    \left[ \alpha_{q}^{r+}(\omega) C^>(\omega) +
    \alpha_{q}^{r-}(\omega) C^<(\omega) \right] \, .
\end{eqnarray}
They depend on the rate functions for charge and heat transport, respectively,
\begin{eqnarray}
\!\!\! \!\! \alpha^{r\pm}_e(\omega)\!\! &=&\!\!  \alpha_0^r \int^\infty_{-\infty} \!\!\!dE\,
  f^\pm_r(E+\omega)f^\mp(E)
\\
\!\!\!\!\!  \alpha^{r\pm}_q(\omega) \!\! &=& \!\! \alpha_0^r \int^\infty_{-\infty}  \!\!\!dE \,
  (E+\omega-\mu_r) f^\pm_r(E+\omega) f^\mp(E)  ,
\end{eqnarray}
as well as on the correlation functions $C^{\gtrless}(\omega)$ for the island charge.
The Fermi function $f$ is denoted by $f^+$, while $f^-=1-f$.
Applied temperature or voltage differences, $\Delta T = T_L - T_R$ and
$\Delta V=V_L - V_R$, are accounted for by evaluating $f_r^\pm(E+\omega)$ at
temperature $T_r$ and voltage $V_r$, while $f^\mp(E)$ is taken at the island electron temperature $T$.
Finally, we define $\alpha^\pm (\omega) = \sum_r  \alpha^{r\pm}_e (\omega)$.

In the following we concentrate on the linear-response regime.
Furthermore, we assume that the heat current is conserved, i.e., the heat
current entering the island from one lead leaves the island to the other lead
and the heat flux from the island electrons to lattice and substrate can be neglected.
It is convenient to use current conservation $\sum_r I^r_{e/q} = 0$ to write
the current as 
$I_{e/q} = (\alpha_0^R I^L_{e/q} - \alpha_0^L I^R_{e/q} ) / 
(\alpha_0^L + \alpha_0^R)$, and then expand up to linear order in $\Delta V$ 
or $\Delta T$.
Then, only the {\em equilibrium} correlation functions $C^\gtrless(\omega)$,
taken at $\Delta V=0$ and $\Delta T=0$, enter, which are related to the
spectral density $A(\omega)$ for charge excitations on the island by
$C^> (\omega) = - 2\pi i[1-f(\omega)] A(\omega)$ and
$C^< (\omega) = 2\pi i f(\omega) A(\omega)$.

We introduce dimensionless thermoelectric coefficients
$g_V = G_V/G_{\rm as}$, $g_T = -(e/k_B) G_T/G_{\rm as}$,
$m = -(e/k_BT)M/G_{\rm as}$, and $k=(e^2/k_B^2T) K/G_{\rm as}$, where
$G_{\text{as}}=4\pi^2 (e^2/h) \alpha_0^L \alpha_0^R /(\alpha_0^L+\alpha_0^R)$
is the classical charge conductance asymptotically reached in the
high-temperature limit. The Lorenz ratio is $L=(k_B/e)^2 \left[k/g_V - (g_T/g_V)^2\right]$.
We, then, find
\begin{equation}\label{allcoeffs}
  \left(\begin{array}{cc} g_V &  g_T \\ m & k \\ \end{array}\right)
  =
  \int d\omega \frac{\beta \omega/2}{\sinh \beta\omega} A(\omega)
  \left(\begin{array}{cc} 1 & \frac{\beta\omega}{2} \\
    \frac{\beta\omega}{2} & \frac{\pi^2+(\beta\omega)^2}{3} \\
  \end{array}\right)
\end{equation}
with the equilibrium spectral density $A(\omega)$.

{\it Perturbation expansion.} -- We proceed by performing a
systematic perturbation expansion of the spectral density in the
tunnel conductance $\alpha_0$, as was already done in
Refs.~\cite{KoenigSchoellerSchoenPRB98,kubala06}, based on a
diagrammatic real-time technique \cite{SchoellerSchoen94}, to
address charge conductance and thermopower, respectively. To write
down the first two terms of the perturbation expansion we only have
to specify the results for the correlation functions from
Ref.~\cite{KoenigSchoellerSchoenPRB98} for vanishing voltage
and temperature bias.

By sequential tunneling processes only charge excitations at the resonances, $\omega=\Delta_N$, can be accessed.
The zeroth-order contribution, hence, reads
\begin{equation}\label{1st}
  A^{(0)} (\omega) = \sum_N (P_N+P_{N+1}) \delta(\omega - \Delta_N) \, ,
\end{equation}
where $P_N$ are the equilibrium probabilities
to find $N$ charges on the island.
The next-order contribution is given by
$A^{(1)} (\omega) = \sum_{i=1}^3 A^{(1)}_i (\omega)$ with
\begin{widetext}
~\vspace*{-1.1cm}\\
\begin{eqnarray}\label{2nd1}
  A^{(1)}_1 (\omega) &=& \sum\nolimits_N \left[P_N \alpha(\omega) +
    P_{N-1} \alpha^+(\Delta_N+\Delta_{N+1}-\omega)
    + P_{N+1} \alpha^-(\Delta_N+\Delta_{N-1}-\omega) \right]
  \; {\rm Re} \, R_N(\omega)^2
  \\\label{2nd2}
  A^{(1)}_2 (\omega) &=& \sum\nolimits_N \left(P_N + P_{N+1}\right)
  \delta(\omega - \Delta_N)
  \left[ \partial (2\phi_N + \phi_{N+1} + \phi_{N-1})
    - \frac{\phi_{N+1} - \phi_{N-1}}{E_C}
    \right. \nonumber \\ && \left.
     +\beta \sum\nolimits_{N'} P_{N'} \left( \phi_{N'} - \phi_{N'-1} \right)
    - \beta \frac{ P_{N} \left( \phi_{N} - \phi_{N-1} \right)
      + P_{N+1} \left( \phi_{N+1} - \phi_{N} \right)}{P_N + P_{N+1}}
    \right]
  \\\label{2nd3}
  A^{(1)}_3 (\omega) &=& -\sum\nolimits_N \left(P_N + P_{N+1}\right)
  \delta'(\omega - \Delta_N)
  \left[ 2 \phi_N - \phi_{N+1} - \phi_{N-1} \right] \, .
\end{eqnarray}
Here we used the abbreviations $ R_N(\omega) = 1/(\omega - \Delta_N
+ i 0^+) - 1/(\omega - \Delta_{N-1} + i 0^+)$ as well as $\phi_N =
\alpha_0 \Delta_N {\rm Re} \, \Psi (i \beta \Delta_N/2\pi)$, where
$\Psi(x)$ is the digamma function, and $\partial \phi_N$ is a short
notation for $\partial \phi_N / \partial \Delta_N$. 
\end{widetext}
The first term
describes two-electron cotunneling that describes the charge
excitations away from resonance, $\omega \neq \Delta_N$. The second
and third term are contributions on resonance, $\omega = \Delta_N$,
that, in the low-temperature regime, are identified with corrections
to sequential tunneling due to renormalization of the
tunnel-coupling strength and the charging-energy gap, respectively.
\begin{figure}[b]
\begin{center}
\includegraphics[width=0.48\textwidth]{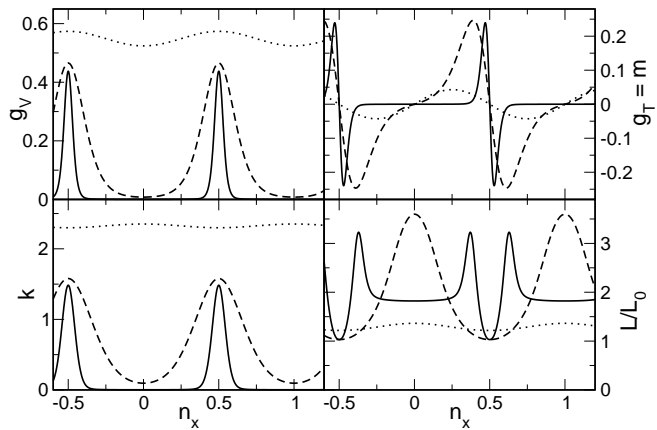}
\caption{
Coulomb oscillations of thermoelectric coefficients and Lorenz ratio. For high temperatures ($k_BT=E_C/2$ \--- dotted line) oscillations are washed out. In the sequential-tunneling regime ($k_BT=E_C/10$ \--- dashed line) the Lorenz ratio is given by Eq.~(\ref{sequential}) around each resonance. For low temperatures ($k_BT=E_C/40$ \--- solid line) the new universal Lorenz ratio $9/5L_0$ is reached in the cotunneling regime.
}\label{fig1}
\end{center}
\end{figure}
\begin{figure}[b]
\begin{center}
\includegraphics[width=0.45\textwidth]{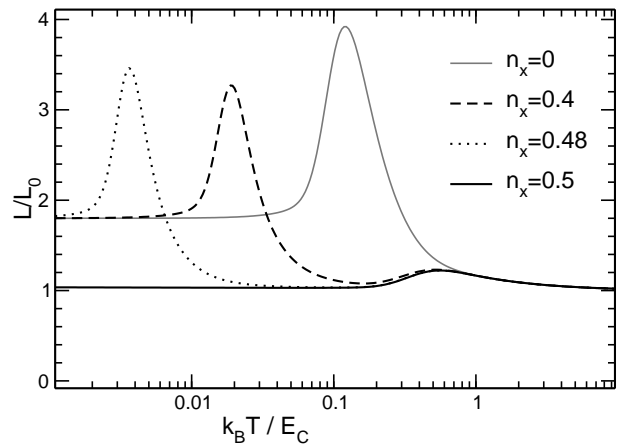}
\caption{
Temperature dependence of Lorenz ratio for different gate voltages. Two maxima separate different tunneling 
regimes. The rise to the maxima starts at $k_BT\simeq E_C$ and 
$k_BT\simeq \Delta_0$, 
respectively. For $n_x=0 \Leftrightarrow \Delta_0=E_C$ the two maxima coincide, whereas at resonance ($n_x=0.5 \Leftrightarrow \Delta_0=0$) the lower maximum is removed to the left.
}\label{fig2}
\end{center}
\end{figure}

{\it Results.} -- Analytical results for the thermoelectric
coefficients can be 
found from Eq.~(\ref{allcoeffs}) 
with Eqs.~(\ref{1st})-(\ref{2nd3}). 
In Fig.~\ref{fig1} we show the
resulting dimensionless thermoelectric coefficients $g_V$, $g_T=m$,
$k$, and the Lorenz ratio $L$ normalized by $L_0$, as a function of
gate voltage for various temperatures; in Fig.~\ref{fig2} the
temperature dependence of $L/L_0$ for various gate voltages. The
tunnel coupling is chosen as $\alpha_0^{L/R}=0.01$. 
Temperature and
gate-voltage dependence of the Lorenz ratio can be elucidated by deriving analytical
expressions for various limits:

(i) In the high-temperature regime, $\beta E_C \ll 1$, Coulomb oscillations are
washed out, i.e., there is no gate-voltage dependence.
To calculate corrections to Wiedemann-Franz law in this regime, we expand the
gate-voltage average of all thermoelectric coefficients in powers up to
$(\beta E_C)^2$ to find
\begin{equation}
  \frac{L}{L_0} = 1 + \frac{2}{\pi^2} \beta E_C
  - \frac{1+24\alpha_0}{3\pi^2} (\beta E_C)^2 \, .
\end{equation}
Deviations from Wiedemann-Franz law are visible before Coulomb
oscillations set
in, see Fig.~\ref{fig2},
where for $k_BT\gtrsim E_C$ the curves coincide for all gate voltages while $L>L_0$.

(ii) In the on-resonance low-temperature regime, $\beta E_C \gg 1$ but
$\beta \Delta_N \ll 1$ for one $N$ (say $N=0$), transport is dominated by sequential
tunneling and only the charge states $0$ and $1$ occur.
The sequential-tunneling contribution then yields 
\begin{equation}\label{sequential}
  \frac{L}{L_0} = 1 + (\beta\Delta_0)^2/(2\pi)^2 \, ,
\end{equation}
in agreement with Ref.~\cite{tsaousidou}.
The Wiedemann-Franz law is only fulfilled for a vanishing
charging-energy gap, $\Delta_0=0$, with corrections quadratic in
$\beta\Delta_0$ away from resonance. These corrections indicate,
that the contribution of each transported particle to the heat
current scales with the charging-energy gap $\Delta_0$ instead of
temperature $k_BT$ as in bulk transport.

Higher-order corrections in 
$\alpha_0$ 
lead to an increase of the Lorenz ratio.
For $\Delta_0=0$ we find
\begin{equation}
  \frac{L}{L_0} = \frac{1 + 4\alpha_0/3 -2\alpha_0 \left[ \gamma + \ln (\beta E_C/\pi) \right]}
    {1-2\alpha_0 \left[ \gamma + \ln (\beta E_C/\pi) \right]} \, ,
\end{equation}
with Euler's constant $\gamma = 0.577 \ldots$.
The terms logarithmic in temperature are associated with the renormalization
of the tunnel-coupling strength, that enter both $g_V$ and $k$ in the same way,
so that it affects the Lorenz ratio only weakly.

(iii) In the off-resonance low-temperature regime, $\beta \Delta_N \gg 1$ for
all $N$, transport is dominated by cotunneling.
Expanding the thermoelectric coefficients up to quadratic order in temperature,
and taking into account only $\Delta_0$ and $\Delta_{-1}$ as the two lowest
excitation energies, we find
$g_V = \alpha_0 (8\pi^2/3) E_C^2/(\beta \Delta_0 \Delta_{-1})^2$,
$g_T=m=0$, and $k = 3\pi^2/5 g_V$.
Because of weak energy dependence of the cotunneling scattering rate, proportionality between charge and heat conductance is recovered, however, with a different prefactor, as
\begin{equation}
  \frac{L}{L_0} = \frac{9}{5} \, .
\end{equation}
\begin{figure}[t]
\begin{center}
\includegraphics[width=\columnwidth]{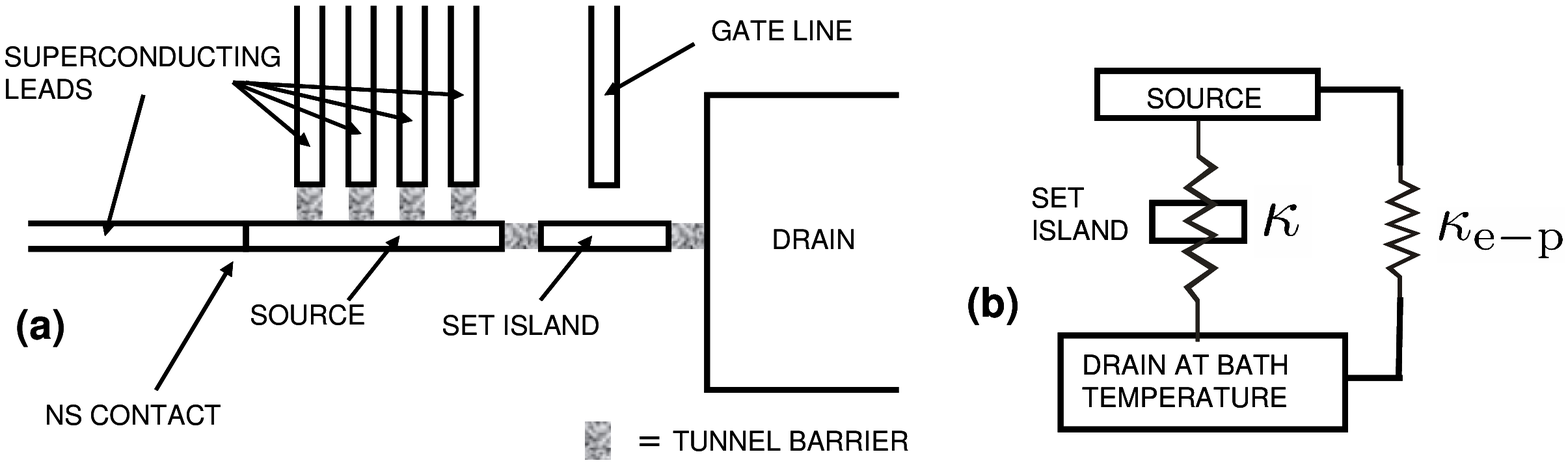} \caption{A possible measurement scheme. In
(a) the SET transistor is depicted in a configuration where the
drain is thermalized at the bath temperature and the source is a
small conductor whose temperature can be varied. In (b) we show a
simple thermal model of the device.
}\label{fig3}
\end{center}
\end{figure}
{\it Possible experimental realization.} -- A possible experimental
configuration to measure the predicted effects is shown in
Fig.~\ref{fig3} (a). The whole structure can be fabricated by
standard nano-lithography and thin film metal deposition and with
oxidized tunnel barriers in the SET. The normal metal source of the
SET is connected to superconducting leads, some of which are tunnel
coupled normal metal-insulator-superconductor (NIS) junctions and the rest are direct normal-metal-superconductor junctions (NS).
Both the NIS junctions and the NS Andreev mirrors provide thermal
isolation of the source \cite{giazotto06}. Another role of the
direct NS contacts is to suppress the charging energy of the source
electrode while keeping it as small as possible to avoid coupling to
phonons \cite{saira07}. The NIS junctions are used for sensitive
thermometry and for heating the source; in proper bias range they
can also cool it \cite{giazotto06}. In contrast to  source electrode 
and SET island the drain is made large (wide
and thick) to secure thermalization at the bath temperature. The NS
contacts are used for characterizing the SET electrically by passing
current and measuring voltage across. 
The structure appears rather simple; we note, however, that combining a
normal metal SET with superconducting probes requires use of
non-conventional material combinations. Figure~\ref{fig3} (b) shows
the relevant thermal model of the proposed setup. When applying
either positive or negative heat flux into the source by a voltage
on the NIS contacts, the dominant energy relaxation mechanism should
be that discussed in this letter, i.e., electronic thermal
conductance $\kappa$ through the SET transistor. 
This requires
$\kappa_{\rm e-p} \ll \kappa$ for a small temperature bias of the
source with respect to the drain at the bath temperature $T_0$. On
the left hand side of the inequality $\kappa_{\rm e-p} =5\Sigma
\mathcal{V} T_0^4$ is the linearized thermal conductance by
electron-phonon coupling from source to the lattice. Here $\Sigma
\simeq 1\cdot 10^9$ WK$^{-5}$m$^{-3}$ is a material specific
constant \cite{giazotto06} and $\mathcal{V}$ is the volume of the
source, which can realistically be made as small as $1\cdot
10^{-21}$m$^{3}$ with a small number of probes attached. A similar 
argument applies to the choice of the size of the SET island. The
inequality between the two heat conductances is satisfied at a
realistic temperature of $T_0 = 30$ mK, in particular near the
resonance; deep in the cotunneling regime one may need to resort to the very different temperature and gate-voltage
dependencies to distinguish contributions of $\kappa$ from that of
$\kappa_{\rm e-p}$.

{\it Conclusions.} --
We have theoretically investigated the influence of Coulomb interaction on thermoelectric transport
coefficients for a metallic single-electron transistor and proposed a measurement scheme for experimental verification.
We found strong violation of the Wiedemann-Franz law. For sequential tunneling the Wiedemann-Franz ratio depends quadratically on the charging-energy gap: the Wiedemann-Franz law is fulfilled only at the resonances, where the charging-energy gap vanishes. In the cotunneling regime the Lorenz ratio takes a new universal value of 9/5 of the Lorenz number.

We acknowledge discussions with T.~Heikkil\"a and thank for the hospitality of the Lewiner Institute for Theoretical Physics at the Technion, Haifa, Israel where this work was initiated.
\vspace*{-0.8cm}\\


\begin{thebibliography}{99}
\bibitem{discrete-dots-exp}
A.S.~Dzurak {\it et al.}, Solid State Commun.~{\bf 87}, 1145 (1993);
A.S.~Dzurak {\it et al.}, Phys.~Rev.~B {\bf 55}, R10197 (1997);
R.~Scheibner {\it et al.}, Phys.~Rev.~Lett.~{\bf 95}, 176602 (2005).

\bibitem{chaotic-dots-exp}
S.~M\"oller {\it et al.}, Phys.~Rev.~Lett.~{\bf 81}, 5197 (1998);
S.F.~Godijn {\it et al.}, Phys.~Rev.~Lett.~{\bf 82}, 2927 (1999).

\bibitem{SmallPRL03}
J.P.~Small, K.M.~Perez, and P.~Kim, Phys.~Rev.~Lett.~{\bf 91}, 256801 (2003); 
P.~Reddy {\it et al.}, Science {\bf 315}, 1568 (2007).

\bibitem{StaringEPL93}
A.A.M.~Staring {\it et al.}, Europhys.~Lett.~{\bf 22}, 57 (1993).

\bibitem{thermopower-theory}
C.W.J.~Beenakker and A.A.M.~Staring, Phys.~Rev.~B {\bf 46}, 9667 (1992);
Ya.M.~Blanter {\it et al.}, {\it{ibid.~}}{\bf{55}}, 4069 (1997);
M.~Turek and K.A.~Matveev, {\it{ibid.~}}{\bf{65}}, 115332 (2002);
K.A.~Matveev and A.V.~Andreev, {\it{ibid.~}}{\bf{66}}, 045301 (2002);
J.~Koch {\it et al.}, {\it{ibid.~}}{\bf{70}}, 195107 (2004);
M.~Turek, J.~Siewert, and K.~Richter, {\it{ibid.~}}{\bf{71}}, 220503(R) (2005).

\bibitem{kubala06}
B.~Kubala and J.~K\"onig, Phys.~Rev.~B {\bf 73}, 195316 (2006).

\bibitem{BoeseEPL01}
D.~Boese and R.~Fazio, Europhys.~Lett.~{\bf 56}, 576 (2001).

\bibitem{tsaousidou}
X.~Zianni, Phys.~Rev.~B {\bf 75}, 045344 (2007);
M.~Tsaousidou and G.T.~Triberis, AIP Conf.~Proc.~893, 801 (2007); cond-mat/0605286 (unpublished).

\bibitem{Kondo-dot}
B.~Dong and X.L.~Lei, J.~Phys.: Condens.~Matter {\bf 14}, 11747 (2002);
M.~Krawiec and K.I.~Wysokinski, Phys.~Rev.~B {\bf 73}, 075307 (2006).

\bibitem{open-dot}
M.G.~Vavilov and A.D.~Stone, Phys.~Rev.~B {\bf 72}, 205107 (2005);
Y.~Ahmadian, G.~Catelani, and I.L.~Aleiner, Phys.~Rev.~B {\bf 72}, 245315
(2005).

\bibitem{beloborodov05}
I.S.~Beloborodov {\it et al.}, Europhys.~Lett.~{\bf 69}, 435 (2005).

\bibitem{many_channel}
The index $n$ enumerates the transverse channels, which includes
the spin, while the wave vectors $k$ and $q$ label the states of the
electrons within each channel.
In the following, we assume the many-channel limit.

\bibitem{KoenigSchoellerSchoenPRB98}
J.~K\"onig, H.~Schoeller, and G.~Sch\"on, Phys.~Rev.~B~{\bf 58}, 7882 (1998).

\bibitem{SchoellerSchoen94}
H.~Schoeller and G.~Sch\"on, Phys.~Rev.~B {\bf 50}, 18436 (1994);
J.~K\"onig, H.~Schoeller, and G.~Sch\"on, Europhys.~Lett.~{\bf 31}, 31 (1995).

\bibitem{giazotto06}
F.~Giazotto \textit{et al}., Rev.~Mod.~Phys.~{\bf 78}, 217 (2006).

\bibitem{saira07}
O.-P.~Saira \textit{et al}., Phys.~Rev.~Lett.~{\bf 99}, 027203
(2007).

\end{thebibliography}
\end{document}